\documentclass[12pt]{article}
\usepackage[]{latexsym}
\begin{document}

\title{{\bf{Delta baryons in the separation geometry model.}}
\author{G Filewood\\
Research Centre for High Energy Physics,\\
School of Physics,
University of Melbourne,\\ 
Parkville, Victoria 3052 Australia.}
\date{\today}
\abstract
Extension of the separation
 geometry model of baryon
structure from physics/0109024 and
hep-ph/0201270 to the spin 3/2 
Delta baryons. 
Theoretically derived masses in MeV; 
$\Delta^{++}=1240.0,\;\Delta^{-}=1243.4,\;
\Delta^0=1233.9,\;\Delta^+=1232.6$ 
with the first one differing
considerably from the quoted empirical value.
Mass difference values are discussed.}

%hep-ph/0212328 passwd j9y6i
\maketitle

\section{Motivation}
In two previous papers \cite{SG} the background ideas
and methodologies of separation geometry were described in some
detail. The purpose of this paper is to outline a 
simple method of
extension of the calculation technique to the
spin 3/2  baryon resonances
${\Delta}^-,\;{\Delta}^0,\;{\Delta}^+,\;{\Delta}^{++}$ as an extension of 
the proton and neutron mass calculations presented previously
and to test the result against the recent calculation
of Capstick et. al {\cite{Capstick}} for the mass differences of these
objects. 

\section{Brief background concepts.}

Separation geometry approaches the issue of physical structure
from quite a different perspective to standard QFT. Instead of 
superimposing fields satisfying local gauge invariance on a 
background four-dimensional space-time continuum, separation 
geometry works with models of particles as geometry  based on explicitly 
{\it{local gauge dependent}} dimensional decomposition of the 
four-dimensional space-time continuum. This decomposition is 
well defined  and is isomorphic to the cardinality structure of
the real number continuum; i.e. it
presupposes that space-time is a real
continuum. The dimensional decomposition reduces 
fields to a local-gauge dependent form which is found to be 
suitable for the calculation of masses of fundamental fermions
and the vector gauge bosons; in large measure because the 
problem of infinities associated with renormalisation of
QFT's and running gauge couplings 
are eliminated in such a local gauge 
dependent approach; which fixes the coupling scale as a result
of fixing the local-gauge 
(local phase) of the objets but in a way which
allows for a natural transition to local-gauge independent
geometry in the continuum limit and in which
the geometry representation theory is independent of the
actual local `gauge' selected so that the calculated
masses are likewise 
ultimately independent of local phase information.
 Separation geometry appears to be 
complementary to QFT; it has strengths in precisely the areas that
QFT/standard model  is weak 
(fermion masses, free parameters, no logical underpinning
of the origin of fermion generation structure, raison d'etre for 
gauge structure etc.) but is weak in precisely the areas that QFT 
is strong (calculation of dynamical parameters; decay times, cross-sections
etc).

The geometries that define local-gauge {\it{dependent}} dimensionally
decomposed  fields were called {\it{affine geometries}} in the previous papers
 and  their properties were defined and studied. 
The peculiar property of affine geometry is
that, whilst it allows us to define a quantum object
explicitly in terms of a local phase, it renders
that local phase unmeasurable. The geometric 
invariants of these objects 
 are assumed to manifest as physical observables in the continuum
limit; intrinsic spin, charge, mass etc. For each invariant there 
is a physical observable and 
for each  intrinsic quantum physical observable there is 
a geometric invariant (properties such as momentum of
a lepton are `extrinsic' variables and independent of the geometry).
The local phase (gauge) of the object is never an
observable (the structure of the continuum in the
theory prevents this).
The geometries define bounded spaces which
in the limit of continuous geometry generators 
must, because of the geometrical construction 
of the theory, define compact group 
symmetries with the exception of the foundation geometry (which
is a one-dimensional interval whose length is the gauge 
property) which evolves in the continuum limit to a non-compact 
symmetry associated with translations in space. It is 
an unproven supposition that these local-gauge-dependent
features lead, with dimensional `reconstitution', to local-gauge
{\it{independent}}, i.e. physical, fields although significant
data has been retrodicted (along with some precision predictions)
which lend  support to this supposition. 

As should be expected for a theory of fundamental structure
the theory has extreme economy; there are essentially
only two affine geometries of interest. These are the
affine  cubic and tetrahedral geometries (and their associated
sub-geometries).  In
the calculation process the cube is reduced to tetrahedral equivalent
sub-groups so  the geometry of the tetrahedon and its' associated 
sub-geometries, along with the geometry of
 the real number continuum,  constitute the essential
geometric elements of the theory. All the physical 
observable structure 
is abstracted from just the geometry, ultimately, of a 
tetrahedron in various incarnations. 

\section{Calculation algorithm.}
A set of rules has been developed {\cite{SG}}
which makes the 
calculation process for the mass of particles 
relatively simple. These rules have been derived from
the discrete version of QCD which is a  consequence of 
the embedding of tetrahedral affine symmetry
into cubic affine symmetry. Discretised QCD
has an explicitly gauge-dependent discrete
symmetry in colour space but has many
features that resemble standard QCD.

The rest-mass calculation of a hadron in separation geometry
 is handled in pieces. Each `piece',
with an appropriate non-perturbative radiative 
correction,  of mass is then added
to give a total mass. The pieces are;

1; Constituent quark mass. This is due to the 
energy-momentum of the current quarks and is
represented as the matrix order (the cardinality or
number of matrix
 elements in a set) which, in
the discrete version of QCD, is represented as the 
number of tetrahedral-equivalent matrix units in 
a six-tetrahedral-component 
vector object called the 'particle vector'. This is 
an irreducible representation of the symmetry
(whilst it is probably not an irreducible 
representation of discrete
$SU(3)_c$ - none other is known -
it is  the minimum required to express the 
full $S_8$ cubic 
permutation symmetry as a tetrahedral embedding;
$T_r$ is irreducible so the $T_c$ SU(3) embedding
is irreducible and the result follows).

2; Gluon energy; found in a similar way by adding up
the matrix order of the analogous representations of the
gluons which couple to the particle vector.

3; Current quark intrinsic mass; this is also expressed in
terms of tetrahedral units and represents the 
effective rest-mass of the individual quarks. This is 
calculated from matrix `operators', also formed as 
six-tetrahedral-component objects, 
which couple to the
particle vector to describe the state present. 

4; Current quark separation energy; rather like a potential
energy of separation of the current quarks 
due to the strong interaction at the energy
scale of the calculation which is fixed by the symmetry.
These are termed U(1) components in the text because
there is the suggestion that they are related to 
a discrete U(1) symmetry.
(The electromagnetic potential energy of separation of the 
current quarks is automatically incorporated in the 
 the current-quark
 `operators' structure and associated 
radiative correction - which are non-perturbative and
governed by a semi-empirically determined ansatz; see below).

The details 
in the case of the nucleons are covered in the mentioned papers
\cite{SG}. One identifies the
the order of the various components and then multiples
by the matrix order of the tetrahedral group(s) which
is either 22 or 24 elements depending on 
whether the two group generators are acting as
massless intrinsic fermion-spin generators (22 elements) or not
(in which case you have 24 massive elements);
and then one adds them all up. For second
and third generation quarks, scalar  components arising explicitly from
the Higgs field must be added to the current quark masses calculated
but these are not required for the first generation
quarks (which do not
acquire scalar components in the
separation geometry model; at least not explicitly
- analogous to treating the mass as (?dynamical)
in origin independent of the Higgs field).

 All components, with the exception of 4, acquire
a simple multiplicative 
radiative correction of the form ${\cal{R}}=
(1+\alpha_{q^2=m_e^2}
+G_f)$  where $\alpha$ is the electro-magnetic
coupling strength 
and  $m_e$ is $\approx$ the electron rest mass (which is
roughly equivalent to the mass of a single tetrahedral unit) and
$G_f$ is the weak coupling constant expressed as a dimensionless 
number to represent its' effective strength with respect to 
$\alpha_{em}$ at the low energy scale;
 here of order $10^{-5}$. Here the digit `1'
in ${\cal{R}}$ is 
also functionally the strong coupling constant 
when applied to quarks - the scale of 
$\alpha_s$ is fixed by the tetrahedral symmetry at unity
(this is the great advantage of calculating in 
an explicitly local-gauge dependent discrete environment
where one does not have a running coupling to 
deal with but instead has a fixed point scale;
all mass calculations reduce to the tetrahedral scale
- roughly 0.5MeV - and the radiative correction
is universal across fermion species as we have 
in the discrete scheme quark/lepton unification at
the level of tetrahedral symmetry).
In this sense then, component 4 has a multiplicative 
radiative correction
of ${{\cal{R}}^s}=\alpha_s=1$ when applied to strongly interacting
particles.

After performing the appropriate summation
and applying the non-perturbative 
radiative correction 
the mass of the particle can then be calculated
by, for example 
(and this is usually the simplest way),
  taking the ratio with the electron rest  mass
which in separation geometry is defined
by the order of the tetrahedral $T_r$ group which has
4!=24 elements in its' matrix representation and two generators.
The generators manifest as massless intrinsic spin generators
in the transition to a field theory so that the remaining 22 $T_r$
matrix elements, with radiative correction ${\cal{R}}$,
 defines the electron rest mass;

\begin{equation}
{\cal{R}}.(T_r\;{\mbox{(No. of irrep. matrix elements)}}
\;-T_r\;{\mbox{(generators)}})={\cal{R}}.(4!-2)
{\equiv}
0.5110000\;MeV
\label{massdef}
\end{equation}

 It is then a simple matter
to convert any matrix order expression, $M$ for the mass
of a hadron into MeV;

\[
{\mbox{mass (MeV)}}=
{M\over{{\cal{R}}.22}}.0.511
\] 

\noindent
where M includes any radiative corrections as described.
The multiplicative radiative correction ${\cal{R}}$ 
is a dimensionless number whose value is approximately 1.0073115 and
represents the sum $(1+\alpha^{-1}_{q^2=m_e^2}+G_f)$.
Thus matrix order expressions have the dimension of energy.

\section{Modifications to calculation algorithm for $\Delta$ baryons.}

The delta baryons ${\Delta}^+,\;{\Delta}^{++},\;{\Delta}^0\;\mbox{and}
\;{\Delta}^{-}$
are spin 3/2 fermions with three current quarks;
$I(J^P)={3\over2}({3\over2}^+)$. For mass calculations
of baryons containing only first generation quarks we have the following
mass components to compute;

1. Constituent quark energy.

2. Current quark mass.

3. Gluon energy.

4. Current quark (strong or U(1)) potential terms.

We expect that a shift in spin state will essentially leave 2,3 and 4
unchanged in comparison with the proton and neutron calculations (modulo
adjustments for the different  current quark content 
in individual $\Delta$'s) 
but result in an increase in the value of item 1. 
The simplest ansatz that could be proposed is to increase the effective 
constituent energy by the equivalence of one unit of spin; that is two
units of constituent  quark energy (each unit representing one
half-integer of spin). Since a baryon has
three quarks, this is the same as multiplying the
constituent energy of the nucleon baryon by a factor of 5/3. 
 The actual quark content of the
baryon is carried in the current quark representation - not in the 
constituent `particle vector' representation
 which represents energy above and beyond the current 
quark rest mass due to current quark momentum. This procedure 
seems to work well
for the delta masses.

\section{The calculations.}

We will compute the current quark masses for each of the 
four species first. The ${\Delta}^{++}$ consists of three up
quarks and the current quark representation is;
 
\begin{equation}
\mbox{strong component}=
\left(
\begin{array}{ccc}
\cal{I}&q^*&q^*\\
q^*&\cal{I}&q^*\\
q^*&q^*&\cal{I}\\
\end{array}
\right),
\;\;\;
\mbox{E.M. component}=
\left(
\begin{array}{ccc}
{I}&q^*&q^*\\
q^*&{I}&q^*\\
q^*&q^*&{I}\\
\end{array}
\right)
\label{cq}
\end{equation}

\noindent
and the matrix orders are read off the table;
in the strong component each $q^*$ and each $\cal{I}$ delivers
4! matrix elements and in the E.M. table each $q^*$ gives a (4!-2) and 
each identity a 4! of elements. There are no 
cancellations. This gives 420 matrices. There is a parity
doubling to 840.

To calculate the U(1) components for a baryon we use a 
triangle diagram;
we place one of the current quarks at each vertex and
each line of the triangle represents a potential
energy of separtion. 
%insert epsfig up-up-up.
 Each line 
between two quarks has an energy determined by the quarks at either
end of the line. An up-up bond has 2(4!-2) matrix order, and u-d
line has (4!-2) order and a d-d type line has matrix order 4.(4!-2). We 
sum over the triangle so the ${\Delta}^{++}$ has a U(1) matrix order
of  6.(4!-2) or three up-up bonds.
(These values are derived from the identities of the corresponding
`strong' components of the current quark representation 
coupled to massless  generators with a
${\cal{I}}$ canceling an $I$ so that an up-up interaction
is for example ${\cal{I}}+{\cal{I}}=2$ etc).
 The ${\Delta}^{++}$
total  current quark mass by the 
algorithm is then;

${\Delta}^{++}={\cal{R}}840+6(4!-2)$.
(Note that the U(1) component does not pick up a radiative correction).

For the ${\Delta}^{+}$ and ${\Delta}^{0}$ we have current quark masses
identical the the proton and neutron respectively which have been 
calculated in hep-th/0109024 as; 

${\Delta}^{+}={\cal{R}}564+4(4!-2)$

and;

${\Delta}^{0}={\cal{R}}576+6(4!-2)$

and finally for the ${\Delta}^{-}$ we have three down quarks as per the
chart;

\begin{equation}
\mbox{strong component}=
\left(
\begin{array}{ccc}
{I}&q^*&{I}\\
{I}&{I}&q^*\\
q^*&{I}&{I}\\
\end{array}
\right),
\;\;\;
\mbox{E.M. component}=
\left(
\begin{array}{ccc}
{I}&q&{I}\\
{I}&{I}&q\\
{q}&{I}&{I}\\
\end{array}\right)
\end{equation}

which has the order 15.4! + 3.(4!-2). With parity doubling and
the addition of the U(1) for three down-quarks = 12(4!-2) we 
obtain;

${\Delta}^{-}={\cal{R}}.852+12(4!-2)$.

The glue order for the baryon is easily calculated as 
${\cal{R}}(6.4!)^2$ (this is identical to the value for
the nucleons)
and the constituent quark energy as;

\[ 
{\cal{R}}.{5\over3}(6(4!-2).6.4!)\]

\noindent
(Notice the ${5\over3}$ factor which is the boost to
the constituent energy in the transition from the 
nucleon expression for the constituent mass
 to the $\Delta$ baryons).
An easy calculation then gives the following masses;

${\Delta}^{++}$ =  1240.03 MeV.

${\Delta}^{+}\;\;$  =  1232.61 MeV

${\Delta}^{0}\;\;\;$  =  1233.90 MeV

${\Delta}^{-}\;\;$  =  1243.36 MeV.

\noindent
Note that $\Delta^0-\Delta^+\approx1.3MeV$
and
 $\Delta^{-}-\Delta^{++}\approx3.3MeV$
so  that
$3(\Delta^0-\Delta^+){\approx}{\Delta^{-}-{\Delta}^{++}}$
broadly in agreement with model expectations given by
Jenkins et. al \cite{JL} and 
Capstick et al\cite{Capstick} who predict a value of $\approx1.5MeV$ 
and $\approx4.5Mev$  for these
mass differences. The calculated
mass of the $\Delta^{++}$ in particular
 differs  significantly  from the standard quoted 
empirical value
however;

\[
{\Delta}^{++}=1230.9{\pm}0.3,\;\;\;
{\Delta}^{+}=1234.9{\pm}1.4,\;\;\;
{\Delta}^{0}=1233.6{\pm}0.5.
\]

\noindent
Note that the relation \cite{JL}; 
\begin{equation}
\Delta_3={\Delta^{++}}-{\Delta^{-}}-{3}(
{\Delta^{+}-\Delta^0})=
{{\epsilon^{''}\epsilon^{'}}\over{{\mbox{N}}_c^3}}
\approx
10^{-3}
\label{NSS}
\end{equation}
quoted in \cite{Capstick}
is violated with the derived masses in this study 
as we obtain (changing signs in accordance with 
the mass heirarchy derived);

\[
\Delta_3={\Delta^-}-{\Delta^{++}}-{3}(
{\Delta^0-\Delta^+}){\approx}0.6MeV.
\]
Here two $\epsilon$'s are isopsin violating parameters
for the strong and electromagnetic mass splitting
respectively suggesting that in the model presented
these isospin symmetries are broken.
Interestingly, however, this relation is satisfied exactly
for the strong interaction U(1) components. From 
current quark triangle diagrams one easily obtains;  

\begin{equation}
{\Delta^-_{U(1)}}-{\Delta^{++}_{U(1)}}
=3.({\Delta^0_{U(1)}}-{\Delta^+_{U(1)}})
\label{CC}
\end{equation}
\noindent
and this is exact. From this we might assume that
strong isospin symmetry is preserved. However,  that this
is not apparently the case is seen from the mass hierarchy
conventionally
expected on the isospin scale;
\[
\Delta^{-}(I_3=-{3\over2})>\Delta^{0}(I_3=-{1\over2})
>\Delta^{+}(I_3=+{1\over2})>\Delta^{++}(I_3=+{3\over2})
\]
\noindent
with the masses decreasing with increasingly positive isospin 
values. The separation geometry calculation suggest
that there  is a mass difference between the $\Delta$
resonances favouring the negative isospin values but
that there is also a mass scale that is dependent on
the absolute value of the isospin 
and not dependent upon sign so that $|I_3=\pm{3\over2}|$
states are  more massive than $|I_3=\pm{1\over2}|$ states.

If we ignore the strong-interaction
 U(1) components completely
(i.e. remove them from the mass calculation)
separation geometry gives another  exact mass relation
between the mass difference of the $|I_3=\pm{3\over2}|$
$\Delta$'s and the $|I_3=\pm{1\over2}|$ states;

\begin{equation}
\Delta^{-}-\Delta^{++}=\Delta^{0}-\Delta^{+}
\label{SS}
\end{equation}

\noindent
The existence of exact relations eq.(\ref{CC}) and 
eq.(\ref{SS}) suggests that there are symmetries
related to isospin in the separation geometry
model of the Delta resonances which are exactly
preserved for the strong interaction   eq.(\ref{CC})
and the electro-magnetic interaction  eq.(\ref{SS})
but that the relationship is more complex than is 
conventionally represented.

The most important way the
separation geometry model of current quarks differs
from the standard model is in terms of the
identities (the ${\cal{I}}$'s and the $I$'s) in the 
current quark operator structure. These have no analogue in
standard model. Note that if these identities are 
treated as scalars (although it may be that they
should actually be treated as spin 1 rather than scalar
which
amounts to a global gauge redefinition  of the
intrinsic spin of the quarks uniformly and  presumably 
no observable consequence?)
 then the up and down quarks
become super-partner particles as composite
scalar / fermion fusions with the scalar (?spin 1) part
representing the `holes' in the charge topology -
for example the `missing' 1/3rd charge in the up
quark is represented by the ${\cal{I}}$ piece in the operator
which carries no electro-magnetic charge but is physically
realised in terms of the strong U(1) components and also
appears in the mass sum of the em charged
current quark operator where it
is `camouflaged' - which is to say its' mass
is blended into the q and q* operators  (recognisable
from the appearance of an ${\cal{R}}$
radiative correction) and presumably
not independently measurable or observable. 

However, the current quark bosonic identity contributions to 
eq.(\ref{SS})  cancel out and mass 
differences here are purely based on the difference 
in massless generator content of the current quark q
and q* operators from the EM components;
i.e. the fermionic electromagnetic generators. 
Both the left and right hand sides of eq.(\ref{SS})
give 12 matrix units which geometrically is the 
number of generators needed to cover (`charge') the
 surface of the cubic
quark / baryon analogue (they have equivalent topology);
 two per square surface (one `square' is one 
$T_r$ unit equivalent) and is the analogue of
a unit of electromagnetic charge
on a cubic baryon. The proton and neutron
have exactly this form (ignoring the strong U(1) 
contributions) the neutron mass is given as
${\cal{R}}8!$ and the proton as ${\cal{R}}(8!-12)$ and similarly
the absolute charge difference between the 
two $I_3=\pm{3\over2}$
$\Delta$'s is one unit of charge topology or 12 matrix units
with the R.H.S. of eq.(\ref{SS}) being 
identical to the proton and neutron E.M. 
mass difference.

  Note that
relation eq.(\ref{SS}) does not represent the {\it{physical}}
$\Delta$ states but states stripped of current-quark
strong interaction potential energy terms.

Lastly note that the precision prediction of the 
$\Delta^{-}$ mass is testable as this object has yet
to have its' mass identified empirically. It would be
interesting to have further measurements of the
$\Delta^{++}$ mass also.

\end{document}